\documentclass[12pt]{article}

\usepackage[a4paper,margin=1in]{geometry}
\usepackage{graphicx}
\usepackage{multirow}
\usepackage{booktabs}
\usepackage{url}
\usepackage{hyperref}
\usepackage{amsmath}
\usepackage{setspace}
\usepackage{lineno}
\usepackage{natbib}
\usepackage{caption}
\usepackage{tabularx}
\usepackage{array}
\usepackage{float}
\usepackage[section]{placeins}
\usepackage{authblk}

\hypersetup{
    colorlinks=true,
    linkcolor=blue,
    citecolor=blue,
    urlcolor=blue
}

\title{Context-Aware Workflow Decomposition for Automated Mobile UI Annotation Using Multimodal Large Language Models}

\author[1]{Athar Parvez\thanks{Corresponding author. Email: \texttt{g202393830@kfupm.edu.sa}}}
\author[1]{Muhammad Jawad Mufti}
\author[2]{Muqaddas Gull}
\author[1]{Omar Hammad}

\affil[1]{Information and Computer Science Department, King Fahd University of Petroleum and Minerals, Dhahran 31261, Saudi Arabia}

\affil[2]{SDAIA--KFUPM Joint Research Center for Artificial Intelligence, King Fahd University of Petroleum and Minerals, Dhahran 31261, Saudi Arabia}

\date{}

\begin{document}

\maketitle

\begin{abstract}
Accurate mobile user interface annotation is important for UI understanding, accessibility tools, automated testing, dataset construction, and GUI agents. However, mobile screens are difficult to annotate because they often contain small, dense, nested, and visually ambiguous elements. Multimodal large language models can help automate this process, but their outputs are sensitive to prompt design and the organization of annotation tasks. This paper studies automated mobile UI annotation from a workflow design perspective, focusing on improving annotation precision. Rather than asking the model to annotate all UI elements in a single step, the task is divided into smaller context-aware stages, allowing related UI elements to be handled with clearer instructions and useful screen context. The proposed pipeline uses structured prompts, schema-constrained JSON outputs, and element-specific annotation instructions. Experiments are conducted on expert-annotated mobile UI screens from the MUIAnno dataset, using eight common UI element types: button, tab, clickable text, card, label, plain text, icon, and image. Four workflow strategies are evaluated: one-step, two-step, four-step, and eight-step annotation. Results show that the two-step workflow achieves the highest precision, while deeper decomposition improves recall but produces more false positives. Additional grouping experiments show that annotation quality depends on both workflow depth and element-class grouping. Overall, careful workflow design can make LLM-based mobile UI annotation more reliable for UI understanding, dataset construction, and GUI agent development.
\end{abstract}

\noindent\textbf{Keywords:} Mobile UI annotation; multimodal large language models; context-aware workflow decomposition; automated UI annotation; GUI grounding; mobile interface understanding

\section{Introduction}
Mobile applications have become one of the main ways people access digital services, complete daily tasks, communicate, learn, shop, manage finances, and interact with information. Behind these interactions is the mobile user interface, which acts as the visible and functional layer between the user and the application. A single mobile screen may contain many different components, including buttons, icons, labels, input fields, cards, tabs, images, and navigation elements. These components are not only visual objects; they also carry functional meaning. A button suggests an action, a tab organizes navigation, a text field expects input, and an icon may represent a command, status, or destination. For this reason, understanding mobile user interfaces has become an important research problem in human-computer interaction, software engineering, accessibility, computer vision, and multimodal artificial intelligence. Recent progress in UI understanding has been strongly supported by large-scale datasets and vision-language models. Early resources such as Rico showed the value of collecting mobile UI screens and structural information at scale for data-driven design and interface analysis \citep{deka_rico_2017}. Later datasets such as MUD and MobileViews further expanded mobile UI data collection by focusing on modern UI styles, larger-scale screen coverage, and screenshot-view hierarchy pairs \citep{feng_mud_2024, gao_mobileviews_2024}. Other work has explored UI representation learning, screen summarization, widget captioning, accessibility labeling, and UI design evaluation \citep{li_screen2vec_2021, wang_screen2words_2021, li_widget_2020, chen_unblind_2020, duan_uicrit_2024}. These studies show that high-quality UI data is essential for building systems that can interpret, describe, search, evaluate, and interact with graphical interfaces.

At the same time, multimodal large language models have changed how UI understanding tasks are approached. Models and systems such as ScreenAI, Pix2Struct, Spotlight, ILuvUI, Ferret-UI, and Ferret-UI 2 have shown that screenshots can be treated as rich multimodal inputs that combine visual layout, textual content, spatial structure, and semantic meaning \citep{baechler_screenai_2024, lee_pix2struct_2023, li_spotlight_2023, jiang_iluvui_2023, you_ferretui_2024, li_ferretui2_2024}. This direction has also become important for GUI agents, where models are expected to ground instructions, identify interface elements, and perform actions in real software environments \citep{xie_osworld_2024, cheng_seeclick_2024, wu_os-atlas_2024, qin_ui-tars_2025, chen_guiworld_2025}. However, these applications depend on reliable element-level understanding. A GUI agent cannot tap the correct region if the target element is poorly localized. An accessibility tool cannot describe an interface properly if small icons or clickable text are missed. An automated testing system cannot interact with a screen reliably if the detected elements contain many false positives. Despite this progress, producing high-quality UI annotations remains difficult. Mobile screens are often dense, compact, and visually heterogeneous. Many elements are small, such as icons, status indicators, and tab items. Others are nested inside larger structures, such as labels inside cards or icons inside buttons. Some elements are visually ambiguous: clickable text may look similar to plain text, tabs may resemble buttons, and cards may be defined more by spacing and layout than by strong visual boundaries. These characteristics make UI annotation different from general object detection. The task requires not only recognizing visible regions, but also understanding the role of each component within the surrounding interface context. Manual expert annotation can produce reliable ground truth, but it is costly and time-consuming. Fully automatic annotation, on the other hand, can scale more easily but often introduces noise. Recent work has begun exploring how large language models can support GUI grounding and automatic annotation, including approaches that generate functionality annotations or use model-based grounding for interface agents \citep{li_autogui_2025, hui_winclick_2025, hui_winspot_2025}. However, directly asking a multimodal model to annotate all elements in a screen at once can lead to several problems. The prompt becomes complex, the model must reason over many element types simultaneously, and visually similar classes may be confused. In such settings, the model may increase coverage by predicting more elements, but this often comes at the cost of precision. For automated dataset construction and model-assisted annotation, precision is especially important because false positives reduce annotation reliability and create additional correction work for human reviewers.

This paper studies automated mobile UI annotation from a workflow-design perspective. 
Rather than asking a multimodal LLM to annotate all UI elements in a single step, we examine whether dividing the task into smaller, context-aware stages can improve annotation reliability. The proposed approach uses structured prompts, schema-constrained JSON outputs, and class-specific instructions to annotate eight common UI element types: button, tab, clickable text, card, label, plain text, icon, and image. 
We compare one-step, two-step, four-step, and eight-step workflows, and also analyze how different groupings of UI element classes affect performance. The results show that moderate decomposition, especially the two-step workflow, provides the best precision by reducing prompt complexity while preserving useful screen context. The findings suggest that workflow design and class grouping are important factors in building reliable LLM-based annotation pipelines for mobile UI understanding \citep{n8n}.

The main contributions of this paper are as follows:

\begin{enumerate}
    \item We propose a context-aware workflow decomposition approach for automated mobile UI annotation using multimodal large language models.

    \item We evaluate one-step, two-step, four-step, and eight-step annotation workflows to study how different levels of task decomposition affect annotation reliability, using precision as the primary evaluation metric.

    \item We examine how grouping UI element classes affects annotation quality and show that meaningful semantic grouping can improve precision more effectively than simply splitting the workflow into more stages.

    \item We analyze the trade-off between prompt complexity, preserved screen context, recall, and false-positive generation, showing that moderate decomposition gives a better reliability balance than annotation that is too fine-grained.
\end{enumerate}

The remainder of this paper is organized as follows. Section~\ref{sec:related_work} reviews related work on mobile UI datasets, multimodal UI understanding, GUI grounding, and automated annotation. Section~\ref{sec:methodology} describes the proposed workflow decomposition approach and experimental setup. Section~\ref{sec:results_discussion}
presents the evaluation results and discusses the effects of workflow decomposition and semantic grouping on annotation quality. Section~\ref{sec:conclusion} concludes the paper and outlines future research directions.

\section{Related Work}
\label{sec:related_work}

Mobile user interface understanding has become an active research area because modern applications are increasingly visual, interactive, and dense. A single mobile screen may contain textual content, icons, navigation elements, cards, images, tabs, and multiple clickable regions. Understanding such screens requires more than recognizing visual objects. It also requires reasoning about layout, hierarchy, interaction, and the functional meaning of each component. Prior work in this area can be broadly viewed through three connected directions: UI datasets and representation learning, multimodal UI understanding, and GUI grounding or automated annotation. These directions provide the foundation for the present study, while also showing why workflow design remains an important but underexplored issue in LLM-based UI annotation.

\subsection{Mobile UI Data and Accessibility}

Early research on mobile UI understanding was strongly shaped by the availability of large-scale datasets. Rico was one of the first major resources to collect mobile application screens together with view hierarchy information, making it possible to study mobile interfaces at scale for design mining, layout analysis, and data-driven UI applications \citep{deka_rico_2017}. Rico demonstrated that mobile screens are not only images, but structured artifacts composed of visual regions, text, layout relationships, and interaction properties. Later datasets expanded this direction by improving scale, modern UI coverage, and data quality. MUD introduced a large-scale and noise-filtered dataset for modern UI modeling, addressing the problem that raw UI data can contain noisy, outdated, or inconsistent interface structures \citep{feng_mud_2024}. MobileViews further contributed large-scale screenshot and view hierarchy pairs for mobile GUI understanding \citep{gao_mobileviews_2024}. More recently, AMEX provided a multi-annotation dataset for mobile GUI agents, reflecting the growing need for richer supervision beyond simple screen-level labels \citep{chai2025amex}. These datasets have made UI understanding more systematic, but they also reveal a practical challenge. View hierarchies are useful, yet they do not always match what a human visually understands from a screen. Some hierarchy nodes may be invisible or redundant, while some visually meaningful regions may not be cleanly represented. In addition, mobile screens often contain nested and visually ambiguous elements, such as labels inside cards, icons inside buttons, and text that may or may not be clickable. This makes annotation difficult because the task requires both visual recognition and semantic interpretation.
Table~\ref{tab:related_datasets_representation} summarizes representative work on UI datasets, representation learning, and accessibility-oriented UI understanding, highlighting how each direction relates to the present study.
Another related direction is semantic grouping in mobile interfaces. \citep{xiao_ui_semantic_group_2024} studied UI semantic component group detection, where visually and semantically related UI elements are grouped together in mobile graphical interfaces .  This is relevant to the present study because the proposed workflow decomposition also assumes that related UI elements should not always be treated independently. Instead, grouping visually or functionally related elements can preserve useful context for annotation.
Several studies have therefore focused on learning better UI representations and descriptions. Screen2Vec proposed semantic embeddings for GUI screens and components, showing that UI elements can be represented in a way that captures their contextual and functional roles \citep{li_screen2vec_2021}. Screen2Words studied automatic mobile UI summarization, where models generate natural-language descriptions of complete screens \citep{wang_screen2words_2021}. Widget Captioning focused on generating natural-language descriptions for individual UI components \citep{li_widget_2020}. Similarly, work on describing UI screenshots in natural language showed that UI understanding involves connecting visual layout with meaningful textual descriptions \citep{leiva_describing_2022}. These studies are important because they move UI analysis beyond low-level detection and toward semantic understanding. Accessibility-related research has also highlighted the importance of accurate component-level interpretation. Many mobile applications contain icons, image buttons, and controls without useful labels, which creates barriers for screen-reader users. \citep{chen_ui_2018} proposed predicting natural-language labels for mobile GUI components using deep learning, showing that models can help improve accessibility when labels are missing or incomplete. More recent work has examined the use of large language models to infer alt-text for UI icons during app development \citep{haque_inferring_2024}. These studies are closely related to automated UI annotation because they require the system to identify meaningful components and infer their roles from surrounding context. UI understanding has also been studied from the perspective of design generation, search, and evaluation. \citep{chen_wireframe-based_2020} proposed translating UI design images into GUI skeletons, showing how visual interface designs can be converted into structured implementation artifacts. Wireframe-based UI design search used image autoencoders to retrieve UI designs from structural visual representations. UICrit introduced a dataset for automated UI design critique, showing that models can be used not only to parse interfaces but also to evaluate their quality \citep{duan_uicrit_2024}. Together, these works show that reliable UI data and annotations are central to many downstream tasks, including accessibility, design search, summarization, testing, and automated interaction.

\begin{table}[h]
\centering
\caption{Representative work on UI datasets, representation learning, and accessibility-oriented understanding.}
\label{tab:related_datasets_representation}
\resizebox{\textwidth}{!}{%
\begin{tabular}{p{3.3cm} p{4.4cm} p{5.8cm}}
\hline
\textbf{Work} & \textbf{Main focus} & \textbf{Relevance to this study} \\
\hline
Rico \citep{deka_rico_2017} & Large-scale mobile UI screens and view hierarchies & Established the importance of structured UI data for data-driven interface research. \\
MUD \citep{feng_mud_2024} & Noise-filtered modern UI dataset & Shows the need for cleaner UI data when training or evaluating UI models. \\
MobileViews \citep{gao_mobileviews_2024} & Large-scale mobile GUI dataset & Supports research on screenshot-based and hierarchy-based mobile UI understanding. \\
AMEX \citep{chai2025amex} & Multi-annotation dataset for mobile GUI agents & Reflects the growing need for richer annotations in agent-oriented UI tasks. \\
Screen2Vec \citep{li_screen2vec_2021} & Semantic embeddings of GUI screens and components & Shows that UI components carry semantic and contextual meaning beyond appearance. \\
Screen2Words \citep{wang_screen2words_2021} & Mobile UI summarization & Demonstrates the importance of connecting screen layout with natural-language meaning. \\
Widget Captioning \citep{li_widget_2020} & Component-level description generation & Closely related to element-level UI annotation and semantic labeling. \\
Unblind Your Apps \citep{chen_unblind_2020} & Accessibility labels for GUI components & Highlights the need for accurate interpretation of individual UI elements. \\
\hline
\end{tabular}%
}
\end{table}

\subsection{Multimodal Models for UI Understanding}

With the progress of vision-language models and multimodal large language models, UI understanding has increasingly shifted from task-specific models toward more general models that can read, interpret, and reason over screenshots. Pix2Struct introduced screenshot parsing as a pretraining task for visual language understanding, showing that screenshots provide useful supervision because they combine text, layout, and visual structure \citep{lee_pix2struct_2023}. Although Pix2Struct was not limited to mobile interfaces, it influenced later research that treats UI screens as structured visual-language inputs rather than ordinary images. Spotlight focused directly on mobile UI understanding and proposed using vision-language models with attention to relevant screen regions \citep{li_spotlight_2023}. This was an important step because mobile UI tasks often depend on recognizing small or localized elements. ILuvUI further explored instruction-tuned language-vision modeling of UIs from machine conversations, showing that instruction-style data can help models reason about interface content and user intentions \citep{jiang_iluvui_2023}. ScreenAI extended this line of work by developing a vision-language model for UI and infographics understanding, emphasizing the shared challenge of reading structured visual content that contains layout, text, and graphical elements \citep{baechler_screenai_2024}. More recent work has focused on grounded UI understanding. Ferret-UI studied grounded mobile UI understanding with multimodal LLMs, allowing models to connect language instructions with specific regions of a mobile screen \citep{you_ferretui_2024}. Ferret-UI 2 extended this direction across platforms and aimed for more universal UI understanding \citep{li_ferretui2_2024}. These models are highly relevant to automated annotation because element-level annotation also requires grounding: the model must identify where an element is located and what category it belongs to. However, strong multimodal capability does not automatically solve the annotation problem. In many UI understanding tasks, the model is asked to answer a question or locate a single target. In annotation, the model must identify many elements at once and separate visually similar categories such as buttons, tabs, clickable text, labels, icons, images, and cards. This makes the task sensitive to prompt complexity. If the prompt asks for too many element types together, the model may confuse categories or produce extra detections. If the prompt is split too narrowly, the model may lose useful context between related elements. This tension motivates the workflow decomposition approach studied in this paper. 
Table~\ref{tab:related_multimodal_models} summarizes representative multimodal UI understanding models and highlights how they relate to automated UI annotation.
\begin{table}[h]
\centering
\caption{Representative multimodal UI understanding models and their relation to annotation.}
\label{tab:related_multimodal_models}
\resizebox{\textwidth}{!}{%
\begin{tabular}{p{3.3cm} p{4.4cm} p{5.8cm}}
\hline
\textbf{Work} & \textbf{Main focus} & \textbf{Relation to automated UI annotation} \\
\hline
Pix2Struct \citep{lee_pix2struct_2023} & Screenshot parsing for visual-language pretraining & Treats screenshots as structured inputs containing text and layout. \\
Spotlight \citep{li_spotlight_2023} & Mobile UI understanding with focused vision-language reasoning & Shows the value of region-aware reasoning for mobile UI tasks. \\
ILuvUI \citep{jiang_iluvui_2023} & Instruction-tuned UI vision-language modeling & Demonstrates that instruction-based supervision can improve UI reasoning. \\
ScreenAI \citep{baechler_screenai_2024} & UI and infographics understanding & Supports general visual-language reasoning over structured screen content. \\
Ferret-UI \citep{you_ferretui_2024} & Grounded mobile UI understanding & Connects multimodal LLM reasoning with specific UI regions. \\
Ferret-UI 2 \citep{li_ferretui2_2024} & Universal UI understanding across platforms & Extends grounded UI reasoning beyond a single platform or interface type. \\
\hline
\end{tabular}%
}
\end{table}Another related direction is conversational interaction with mobile UIs. Wang et al. explored how large language models can enable conversational interaction with mobile interfaces, where users express goals in natural language and the system interprets UI content to support interaction \citep{wang_enabling_2023}. This direction further supports the idea that UI understanding requires a combination of visual perception, language understanding, and action-oriented reasoning. For annotation, the same combination is needed, but the output must be structured and reliable enough to be used as data.

\subsection{GUI Grounding Agents and Automated Annotation}

In GUI grounding, a model must connect a user instruction or action goal to a specific location on the screen. OSWorld introduced a benchmark for multimodal agents performing open-ended tasks in real computer environments \citep{xie_osworld_2024}. SeeClick showed that GUI grounding can support advanced visual GUI agents by improving the connection between interface understanding and action execution \citep{cheng_seeclick_2024}. OS-ATLAS proposed a foundation action model for generalist GUI agents, further showing the importance of grounding actions in visual interfaces \citep{wu_os-atlas_2024}. UI-TARS continued this trend by developing native agents for automated GUI interaction \citep{qin_ui-tars_2025}. GUI-World introduced a video benchmark and dataset for GUI-oriented agents, moving the problem from static screens to dynamic interaction sequences \citep{chen_guiworld_2025}. Several recent studies have further expanded GUI grounding benchmarks and techniques. WinClick studied GUI grounding with multimodal LLMs \citep{hui_winclick_2025}, while WinSpot introduced a GUI grounding benchmark for multimodal large language models \citep{hui_winspot_2025}. ScreenSpot-Pro focused on professional high-resolution computer use, showing that grounding becomes even harder when screens are larger, denser, and more complex \citep{li_screenspot-pro_2025}. Aria-UI further studies visual grounding for GUI instructions, showing the importance of accurately localizing target UI elements from interface screenshots \citep{yang_ariaui_2025}. Other work has explored grounding multimodal large language models in GUI environments \citep{lei_grounding_2025}, attention-driven GUI grounding for GUI agents \citep{xu_attention_2025}, and test-time scaling through dynamic visual grounding and modality-aware optimization \citep{wu2025dimo_gui}. More recent work has also studied UI decomposition and synthesis for scaling computer-use grounding \citep{xie_scaling_2025}, as well as diffusion-based and multimodal fusion approaches for GUI grounding \citep{kumbhar_towards_2026, ma_trifuse_2026}. These studies are strongly connected to the present work because both grounding and annotation depend on accurate element localization. However, the goals are different. GUI grounding usually focuses on finding a target element for a given instruction, such as selecting a button, menu item, or input field. Automated UI annotation requires a broader output: the model must detect multiple visible elements, assign each one a category, and avoid producing incorrect regions. This makes annotation especially sensitive to false positives. A grounding model may still be useful if it locates one correct target, but an annotation pipeline becomes less reliable when it produces many extra elements that human reviewers must remove. In this sense, annotation is not only a localization problem, but also a quality-control problem. The output needs to be complete enough to cover the screen, but also clean enough to be useful for dataset construction.

Automatic annotation has therefore become an important research direction. AutoGUI is closely related to this work because it also uses LLMs to reduce the cost of producing GUI supervision \citep{li_autogui_2025}. However, its goal is different: AutoGUI focuses on generating functionality annotations for GUI elements from interaction traces and UI state changes. In contrast, our work focuses on visual element annotation from static screenshots, where the model must detect multiple visible elements, assign each one a category, and avoid producing incorrect regions. Therefore, AutoGUI shows the promise of automatic GUI annotation, but it does not directly study how the annotation workflow should be decomposed when using multimodal LLMs. Scalable video-to-dataset generation has also been explored for cross-platform mobile agents, showing another way to create training data from interaction traces \citep{jang_scalable_2025}. GUIOdyssey introduced a comprehensive dataset for cross-app GUI navigation on mobile devices, further emphasizing the importance of data and annotation for mobile agents \citep{lu_guiodyssey_2025}. In addition, survey work on vision-based mobile app GUI testing shows that reliable visual understanding is essential for automated testing and interaction with mobile apps \citep{yu_vision-based_2025}. Despite these advances, the workflow used to obtain annotations has received less attention than model architecture or benchmark design. In many model-assisted annotation settings, the model is prompted to produce all labels in a single pass. This is simple, but it can overload the model when the screen is dense and the target classes are visually close. On the other hand, separating every element type into an independent step can reduce prompt complexity but may remove the context needed to distinguish related categories. For example, clickable text may be easier to identify when labels and plain text are also considered, while cards may be easier to understand when their internal text and images are visible as part of the same reasoning process. This trade-off is important in mobile UI annotation because interface elements rarely appear in isolation. A button may look similar to a text label, an image may be part of a larger card, and a group of elements may only make sense when interpreted together. Therefore, the way annotation tasks are divided can affect how well the model understands the screen structure. A workflow that is too broad may lead to missed or confused elements, while a workflow that is too narrow may cause the model to ignore useful visual context. This motivates a closer study of workflow decomposition rather than assuming that a single prompting strategy is always sufficient. Table~\ref{tab:related_grounding_annotation} summarizes representative GUI grounding, GUI agent, and automated annotation studies, highlighting how they connect to the present study. The present work builds on these directions but focuses on a different question: how should the annotation process itself be organized when using multimodal LLMs? Rather than treating annotation as a single-shot output, this study examines context-aware workflow decomposition. The goal is not only to detect more elements but to improve the reliability of predicted annotations. This is why precision is treated as a primary metric. Precision, recall, and F1-score are widely used to evaluate classification and detection systems, but they capture different aspects of performance. In automated annotation, precision is especially important because false positives can reduce dataset quality and increase the amount of manual correction required. Overall, prior research shows that mobile UI understanding has advanced rapidly through better datasets, stronger multimodal models, and more capable GUI agents. However, reliable automated annotation still depends on practical design choices around prompts, grouping, context, and workflow structure. This paper addresses that gap by studying one-step, two-step, four-step, and eight-step annotation workflows, as well as different grouping strategies for common UI element types. By focusing on workflow decomposition, the study complements existing work on UI datasets, multimodal UI models, and GUI grounding, while providing practical insight into how LLM-based annotation pipelines can be made more precise and trustworthy.

\begin{table}[h]
\centering
\caption{Representative GUI grounding and automated annotation studies.}
\label{tab:related_grounding_annotation}
\resizebox{\textwidth}{!}{%
\begin{tabular}{p{3.3cm} p{4.4cm} p{5.8cm}}
\hline
\textbf{Work} & \textbf{Main focus} & \textbf{Connection to this study} \\
\hline
SeeClick \citep{cheng_seeclick_2024} & GUI grounding for visual agents & Shows the importance of locating actionable UI regions. \\
OSWorld \citep{xie_osworld_2024} & Open-ended computer-use agent benchmark & Highlights the need for reliable UI understanding in real environments. \\
OS-ATLAS \citep{wu_os-atlas_2024} & Foundation action model for GUI agents & Connects UI perception with action execution. \\
UI-TARS \citep{qin_ui-tars_2025} & Native GUI interaction agents & Demonstrates the growing role of GUI agents in automated interaction. \\
GUI-World \citep{chen_guiworld_2025} & Video benchmark for GUI-oriented agents & Extends GUI reasoning to dynamic screen sequences. \\
WinClick \citep{hui_winclick_2025} & GUI grounding with multimodal LLMs & Relates directly to model-based localization of interface elements. \\
WinSpot \citep{hui_winspot_2025} & GUI grounding benchmark & Provides evaluation resources for multimodal GUI grounding. \\
AutoGUI \citep{li_autogui_2025} & Automatic functionality annotations from LLMs & Closely related to scalable model-assisted annotation. \\
GUIOdyssey \citep{lu_guiodyssey_2025} & Cross-app mobile GUI navigation dataset & Shows the importance of annotated mobile interaction data. \\
\hline
\end{tabular}%
}
\end{table}

\section{Methodology}
\label{sec:methodology}
\begin{figure}[H]
    \centering
    \includegraphics[width=\linewidth]{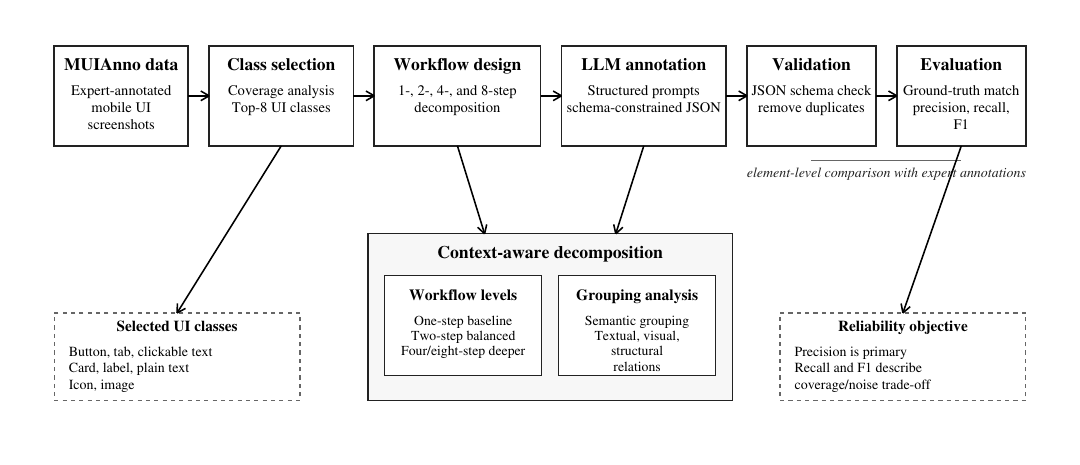}
    \caption{Overview of the proposed methodology for evaluating context-aware workflow decomposition in automated mobile UI annotation. The pipeline starts with expert-annotated MUIAnno screenshots and coverage-based selection of the top eight UI element classes, then applies one-step, two-step, four-step, and eight-step LLM-based annotation workflows. Model outputs are generated using structured prompts and schema-constrained JSON, validated to remove malformed or duplicate predictions, and evaluated against expert ground truth using precision, recall, and F1-score, with precision used as the main reliability measure.}
    \label{fig:methodology_overview}
\end{figure}

This section presents the experimental methodology used to evaluate the proposed context-aware workflow decomposition approach for automated mobile UI annotation. As illustrated in Figure~\ref{fig:methodology_overview}, the methodology begins with expert-annotated mobile UI screenshots from the MUIAnno dataset, followed by the selection of eight common UI element classes. Each screenshot is then processed using multiple LLM-based annotation workflows, including one-step, two-step, four-step, and eight-step strategies. 
The resulting annotations are parsed as structured JSON outputs, validated, and compared against expert ground truth using precision, recall, and F1-score.

\subsection{Research Design and Dataset}

\begin{figure}[H]
    \centering
    \includegraphics[width=0.92\linewidth]{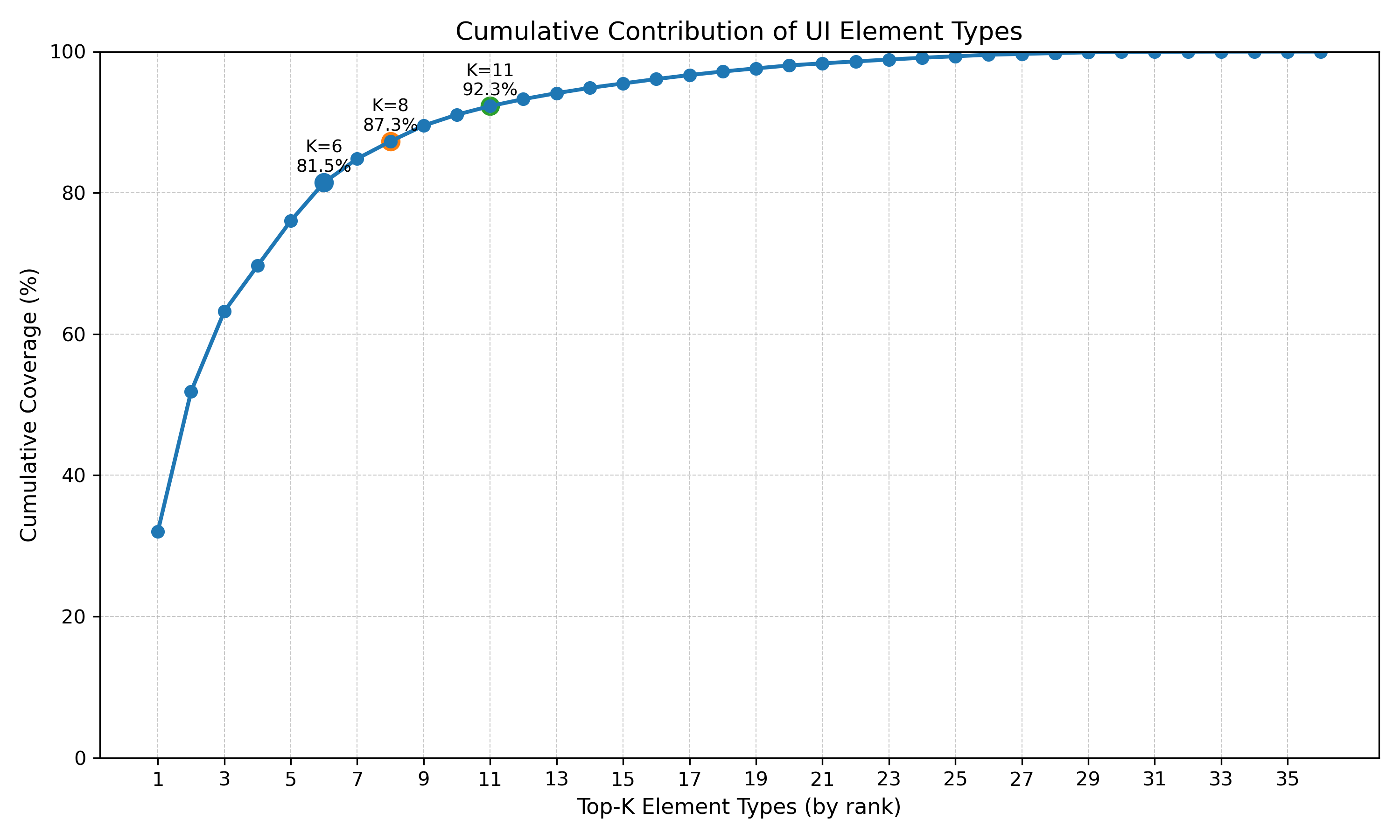}
    \caption{Cumulative coverage analysis of UI element types in the MUIAnno dataset. The selected top eight element classes cover 87.3\% of the annotated elements.}
    \label{fig:element_coverage_analysis}
\end{figure}

This study is designed to evaluate whether the organization of an annotation workflow affects the quality of UI element annotations generated by a multimodal large language model. Instead of treating mobile UI annotation as a single direct prediction task, the study examines annotation as a workflow design problem. The main idea is that a model may behave differently when it is asked to annotate all UI element types at once compared with when the same task is divided into smaller and more focused stages. Therefore, the experiments compare different workflow decomposition strategies under the same dataset, model setting, annotation format, and evaluation procedure. The research design follows a controlled experimental setup. Each mobile UI screenshot is used as the input image, and the model is asked to generate structured annotations for predefined UI element classes. The generated annotations are then compared with expert-provided ground truth annotations. This setup makes it possible to measure how different workflow strategies affect annotation precision, recall, and F1-score. Since the main goal of this work is to produce reliable annotations that can reduce manual correction effort, precision is treated as the primary measure of annotation quality. Recall and F1-score are also reported to understand the trade-off between detecting more elements and avoiding false positives. The experiments are conducted using expert-annotated mobile UI screens from the MUIAnno dataset \citep{parvez2026muianno}. MUIAnno contains mobile interface screenshots with manually verified element-level annotations, making it suitable for evaluating automated annotation methods. The dataset represents realistic mobile application screens, where UI elements often appear in dense layouts and may be visually small, nested, or ambiguous. These characteristics make the dataset appropriate for studying the difficulty of automated UI annotation using multimodal models. Before defining the annotation task, an element coverage analysis was conducted to identify the most frequent UI element types in the dataset. The purpose of this analysis was to focus the experiments on element classes that represent most of the annotation distribution while keeping the workflow manageable. The analysis showed that the top six element types cover 81.5\% of the annotated elements, the top eight element types cover 87.3\%, and the top eleven element types cover 92.3\%. Based on this distribution, this study selects the top eight frequent UI element classes. This choice provides broad coverage of the dataset, nearly 90\% of all annotated elements, while avoiding excessive expansion of the annotation task into many low-frequency classes. Figure~\ref{fig:element_coverage_analysis} illustrates the cumulative coverage of UI element types in the MUIAnno dataset and shows that the selected top eight classes cover 87.3\% of the annotated elements.

The selected eight UI element classes are \textit{button}, \textit{tab}, \textit{clickable text}, \textit{card}, \textit{label}, \textit{plain text}, \textit{icon}, and \textit{image}, as summarized in Table~\ref{tab:selected_ui_classes}. These classes were selected because they are frequent in the dataset and represent the main functional and visual categories found in mobile interfaces. Buttons, tabs, and clickable text represent interactive components. Labels and plain text represent textual components with different semantic roles. Icons and images represent visual components. Cards represent structural containers that may include other elements such as text, icons, or images. Together, these classes provide a balanced setting for evaluating how well a multimodal model can distinguish between interactive, textual, visual, and structural UI elements. \begin{table}[h]
\centering
\caption{Selected UI element classes used in the annotation experiments.}
\label{tab:selected_ui_classes}
\begin{tabularx}{\textwidth}{p{3.8cm} p{4.2cm} X}
\hline
\textbf{UI Element Class} & \textbf{General Category} & \textbf{Reason for Inclusion} \\
\hline
Button & Interactive element & Common action component used for submitting, confirming, opening, or triggering commands. \\
Tab & Navigation element & Represents section-level navigation and may visually overlap with buttons or labels. \\
Clickable text & Interactive text & Important for links, text-based actions, and navigation items that may resemble plain text. \\
Card & Structural container & Groups related content and often contains nested text, icons, buttons, or images. \\
Label & Semantic text & Provides names, titles, descriptions, or identifiers for nearby UI components. \\
Plain text & Non-interactive text & Represents static readable content and helps test distinction from labels and clickable text. \\
Icon & Visual/action element & Often small, symbolic, and visually ambiguous; may appear alone or inside other components. \\
Image & Visual content & Represents non-icon visual regions such as thumbnails, banners, or product/app images. \\
\hline
\end{tabularx}
\end{table} The selected classes also introduce several realistic sources of ambiguity. A tab may look similar to a button when it is presented as a highlighted navigation item. Clickable text may visually resemble plain text if no strong visual cue is present. A label may be confused with plain text depending on its position and role in the screen. Icons may appear alone, inside buttons, or as part of navigation bars. Cards may not always have clear borders and are often defined by spacing, grouping, and layout rather than by a strong visual outline. These cases make the task different from general object detection because the correct class depends not only on appearance, but also on the surrounding interface context. The dataset is used consistently across all workflow strategies. The same screenshots, same target classes, and same ground truth annotations are used for each experiment. This ensures that any difference in performance comes from the annotation workflow rather than from changes in the data. In this way, the methodology isolates the effect of workflow decomposition and class grouping on automated mobile UI annotation quality.

\subsection{Annotation Task Formulation}

The annotation task is defined as an element-level detection and classification problem over mobile UI screenshots. Given a screenshot, the model is expected to identify the visible UI components that belong to the selected target classes and assign each detected component a corresponding class label. Each prediction therefore contains two main parts: the semantic type of the UI element and its spatial location on the screen. The target classes used in this study are \textit{button}, \textit{tab}, \textit{clickable text}, \textit{card}, \textit{label}, \textit{plain text}, \textit{icon}, and \textit{image}.

Formally, for each input screenshot $S$, the annotation output is represented as a set of predicted UI elements:

\begin{equation}
A(S) = \{a_1, a_2, \ldots, a_n\},
\label{eq:annotation-set}
\end{equation}

where each annotation $a_i$ contains an element class $c_i$ and a bounding box $b_i$. The class $c_i$ belongs to one of the eight selected UI element types, while the bounding box $b_i$ defines the position of the element in the screenshot. In this work, bounding boxes are treated as rectangular regions around visible UI elements. Each box is represented using coordinate values that specify the element location and size within the image.

The task is different from general object detection because UI components are not always visually separated objects. Some elements are defined by their function rather than by a strong visual boundary. For example, clickable text may look almost identical to plain text, and a tab may appear as a short text label inside a navigation bar. Similarly, a card may not have a clear border but may still function as a grouped content container because of spacing, alignment, and internal layout. Therefore, the model must rely not only on visual appearance, but also on layout context and semantic cues. The annotation process also requires the model to avoid unnecessary or duplicate predictions. A single UI region should not be repeatedly annotated under different labels unless the dataset definition supports nested elements. For example, a button containing text may be annotated as a button rather than as a separate plain text element if the text functions as part of the button. Similarly, an icon inside a clickable control should be classified according to the target class definition used in the ground truth annotations. This makes consistency important, especially when comparing model-generated annotations with expert annotations. To make the model output machine-readable, each annotation response is requested in a structured JSON format. The use of structured output reduces ambiguity in the generated response and allows the predicted annotations to be parsed automatically. Each returned annotation includes the element type and its bounding box coordinates. Outputs that do not follow the expected structure are treated as invalid and are excluded or corrected according to the validation procedure described in the pipeline. This formulation allows all workflow strategies to be evaluated using the same input screenshots, target classes, and output requirements.

\subsection{Context-Aware Workflow Decomposition}

The main methodological idea of this study is context-aware workflow decomposition. Instead of asking the multimodal model to annotate all selected UI element types in a single prompt, the annotation task is divided into smaller workflow stages. Each stage focuses on a subset of UI element classes. The purpose of this decomposition is to reduce prompt complexity while still preserving enough contextual information for the model to reason about the screen correctly. Mobile UI annotation is highly context-dependent. Many elements can only be classified correctly when their surrounding elements are considered. For example, a short text item at the bottom of a screen may be a tab if it belongs to a navigation bar, clickable text if it performs an action, a label if it names nearby content, or plain text if it only displays information. Likewise, an icon may represent an independent action, part of a tab, or a decorative visual element depending on its position and relation to other components. Because of this, separating UI classes without considering their semantic relationships may reduce the useful context available to the model. At the same time, giving the model too many element types in one prompt can make the task broad and confusing. The model must search for many different types of regions, remember class definitions, distinguish overlapping cases, and return a clean structured output. In dense mobile screens, this can lead to false positives, duplicated detections, or confusion between similar classes. For this reason, decomposition is used as a way to control the scope of each model call. The proposed decomposition strategy is context-aware because the grouping of element classes is not treated as arbitrary. Classes that are visually or semantically related can be placed together so that the model can compare them within the same reasoning step. For example, textual classes such as \textit{clickable text}, \textit{label}, and \textit{plain text} are closely related because they may look similar but differ in function. Visual classes such as \textit{icon} and \textit{image} also require careful distinction because both may appear as non-textual regions. Structural elements such as \textit{card} may depend on the presence of internal text, icons, or images. Therefore, the workflow must balance two needs: simplifying the prompt and preserving the relationships that help the model make correct decisions.

In this study, decomposition is evaluated at different levels. A one-step workflow asks the model to annotate all eight target classes in a single stage. A two-step workflow divides the classes into two broader groups. A four-step workflow further separates the classes into smaller groups. An eight-step workflow annotates each class independently. This design makes it possible to observe whether performance improves when the task is moderately decomposed, and whether excessive decomposition harms precision by removing useful context. The purpose of workflow decomposition is not simply to increase the number of detected elements. In automated annotation, detecting more elements is useful only when the predictions remain reliable. A workflow that increases recall but also produces many false positives may create more work for human reviewers and reduce dataset quality. Therefore, this study evaluates decomposition mainly from the perspective of precision. The goal is to identify a workflow structure that helps the model generate annotations that are correct, consistent, and useful for dataset construction or model-assisted labeling. Overall, context-aware workflow decomposition treats annotation quality as a result of both model capability and task organization. The same multimodal model may produce different outputs depending on how the annotation prompt is structured and how UI element classes are grouped. By studying this effect systematically, the methodology provides insight into how LLM-based UI annotation pipelines can be designed more carefully for practical use.

\subsection{LLM-Based Annotation Pipeline}

The annotation pipeline is designed to make the model-generated outputs consistent, repeatable, and easy to evaluate against the expert annotations. Each mobile UI screenshot is passed to a multimodal large language model together with a structured prompt. The prompt specifies the UI element classes to be annotated in that workflow stage, gives short class-level instructions, and asks the model to return the result in a fixed JSON format. The same general pipeline is used for all workflow strategies so that the comparison between one-step, two-step, four-step, and eight-step annotation remains fair. The pipeline begins with a screenshot from the MUIAnno dataset. For each screenshot, the system prepares the prompt according to the workflow being tested. In the one-step setting, the prompt includes all eight UI element classes. In decomposed settings, the prompt includes only the classes assigned to the current workflow stage. The screenshot and prompt are then submitted to the multimodal model. The model response is expected to contain a list of detected UI elements, where each element includes the predicted class label and its bounding box coordinates. To support automatic processing, the model is instructed to return only structured JSON. This is important because free-form textual responses are difficult to compare with ground truth annotations. A structured response also reduces ambiguity in later stages of parsing and evaluation. Each predicted annotation follows the same basic form: an element type and a rectangular bounding box. The bounding box represents the visible region of the element in the screenshot. Responses that are empty, malformed, or do not follow the expected schema are treated carefully during validation so they do not distort the evaluation. After the model response is received, the JSON output is parsed and checked for validity. The validation step removes predictions with missing fields, invalid class names, or unusable coordinate values. Duplicate predictions are also handled to avoid counting repeated detections as separate correct annotations. This step is necessary because multimodal LLMs may sometimes repeat the same element, produce overlapping boxes for the same object, or include explanatory text even when a structured response is requested. The goal of validation is not to improve the model output manually, but to ensure that the evaluation uses only machine-readable annotations produced by the pipeline. The complete workflow is implemented as an automated pipeline using n8n. The use of n8n makes it possible to run the same annotation process across multiple screenshots and workflow strategies in a consistent manner. It also helps separate the major stages of the experiment: prompt preparation, model call, response collection, JSON parsing, output validation, and result storage. This automation reduces manual intervention and makes the experimental procedure easier to reproduce. Once the predicted annotations are validated, they are compared with the expert ground truth annotations from the dataset. The comparison is performed at the element level. A prediction is considered correct when its predicted class matches the ground truth class and its bounding box sufficiently overlaps with the corresponding ground truth element according to the evaluation protocol. Predictions that do not match any ground truth element are counted as false positives, while ground truth elements that are not detected are counted as false negatives. This allows the same evaluation method to be applied to every workflow strategy.

\subsection{Workflow and Grouping Strategies}

The main experimental variable in this study is the way the eight UI element classes are grouped across annotation stages. The workflow strategies are designed to test whether annotation quality changes when the task is performed in one broad step or divided into smaller steps. All strategies shown in Table~\ref{tab:workflow_strategies} use the same screenshots, same selected element classes, same output format, and same evaluation metrics. The only difference is how the classes are distributed across model calls. The first strategy is the one-step workflow. In this setting, the model receives one prompt for each screenshot and is asked to annotate all eight UI element classes at the same time: \textit{button}, \textit{tab}, \textit{clickable text}, \textit{card}, \textit{label}, \textit{plain text}, \textit{icon}, and \textit{image}. This setting keeps the full screen context in a single model call, but it also gives the model the broadest and most complex task. The one-step workflow is used as the baseline because it represents the most direct way to ask a multimodal model to annotate a screen. The second strategy is the two-step workflow. In this setting, the eight classes are divided into two groups. Each group is annotated in a separate model call, and the outputs from both calls are combined before evaluation. The motivation behind this strategy is that moderate decomposition may reduce prompt complexity without removing too much useful context. The model still reasons over multiple related classes in each step, but it does not need to handle all eight classes at once. The third strategy is the four-step workflow. Here, the selected classes are split into four smaller groups. This provides a more focused prompt for each model call and allows the model to concentrate on fewer element types at a time. However, this strategy also begins to reduce the amount of cross-class context available within each step. For example, if visually similar text classes are separated too much, the model may have less opportunity to compare them directly. The fourth strategy is the eight-step workflow. In this setting, each UI element class is annotated independently. Each model call focuses on only one class. This is the most decomposed strategy and gives the model the simplest prompt in terms of class scope. However, it also removes most of the semantic relationship between classes during prediction. As a result, the model may detect more elements for a single class but may also produce more false positives because it is not comparing that class with related alternatives.

\begin{table}[h]
\centering
\caption{Overview of the evaluated workflow decomposition strategies.}
\label{tab:workflow_strategies}
\begin{tabular}{p{3.0cm} p{3.0cm} p{6.0cm}}
\hline
\textbf{Workflow} & \textbf{Number of Stages} & \textbf{Annotation Setting} \\
\hline
One-step workflow & 1 & All eight UI element classes are annotated together in a single model call. \\
Two-step workflow & 2 & The eight classes are divided into two broader groups and annotated in two separate model calls. \\
Four-step workflow & 4 & The classes are divided into four smaller groups to provide more focused annotation prompts. \\
Eight-step workflow & 8 & Each UI element class is annotated independently in a separate model call. \\
\hline
\end{tabular}
\end{table}

\begin{figure}[H]
    \centering
    \includegraphics[width=\textwidth]{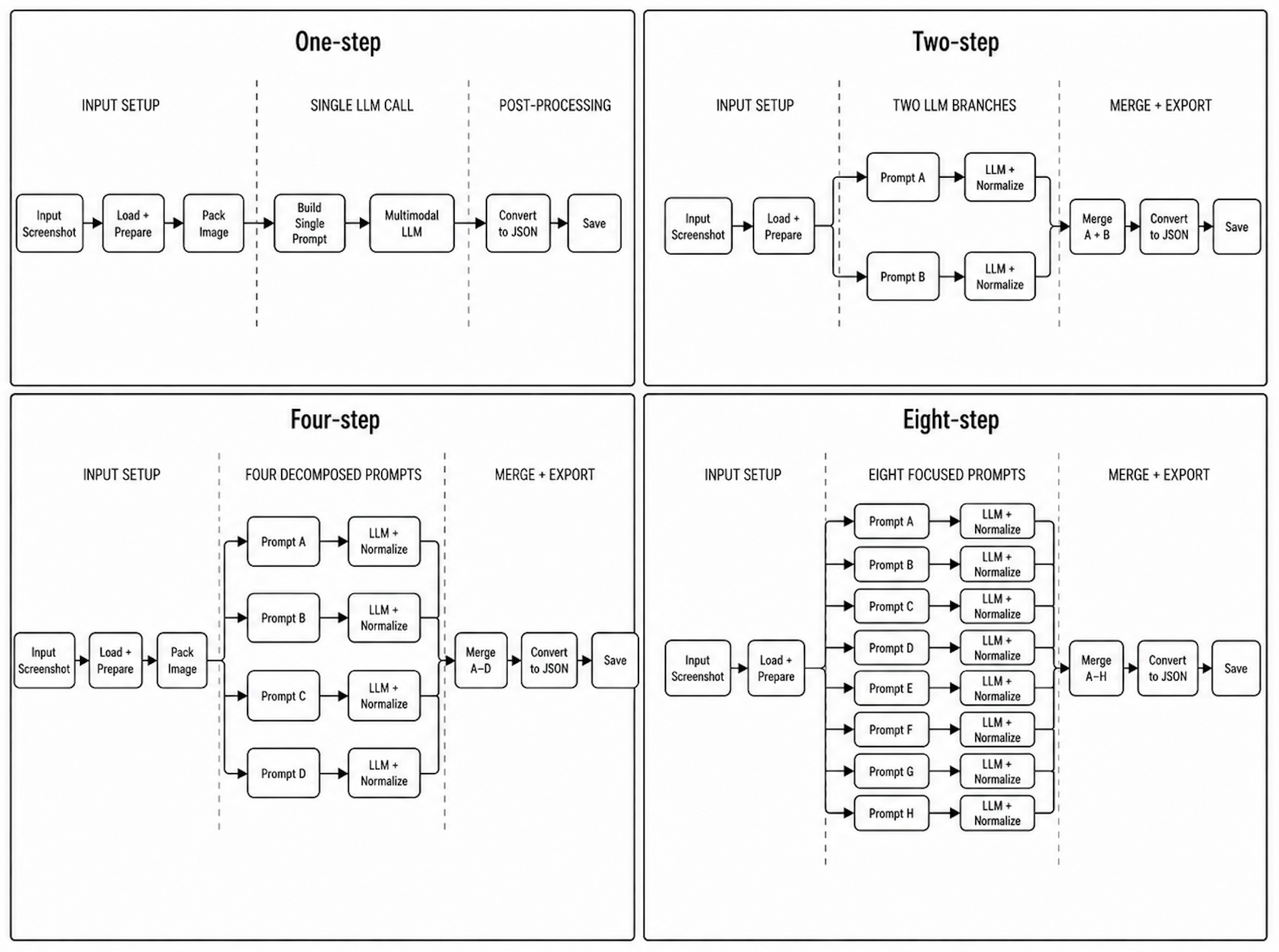}
    \caption{n8n implementation diagrams for the one-step, two-step, four-step, and eight-step workflow decomposition strategies used in the automated UI annotation pipeline.}
    \label{fig:workflow-diagrams}
\end{figure}

In addition to these four workflow levels, the study includes grouping experiments to examine whether the combination of classes inside a workflow stage affects annotation performance. This matters because decomposition is not only about the number of steps: two workflows may have the same number of stages but behave differently depending on which classes are grouped together. The grouping experiments are guided by the semantic and visual relationships between UI elements. Text-related classes, such as \textit{clickable text}, \textit{label}, and \textit{plain text}, may look similar but differ in function. Interactive classes, such as \textit{button}, \textit{tab}, and \textit{clickable text}, represent action or navigation. Visual classes, such as \textit{icon} and \textit{image}, share non-textual appearance but differ in purpose and scale. Structural classes, such as \textit{card}, often depend on nearby text, icons, and images to be recognized correctly. Because of these relationships, some classes may benefit from being annotated together. For example, clickable text may be easier to distinguish from plain text when both are included in the same reasoning step, and cards may be easier to identify when the model can also consider the text and images inside them. However, grouping too many unrelated classes may increase confusion and false detections. The grouping analysis helps determine whether meaningful class combinations can improve precision more effectively than simply increasing the number of workflow stages. After each workflow or grouping strategy is completed, the annotations from all stages are merged into a single prediction set for the screenshot, then validated and evaluated against the same ground truth annotations. To make the implementation of the evaluated workflow strategies clearer, Figure~\ref{fig:workflow-diagrams} shows the n8n-based workflow structure used for the one-step, two-step, four-step, and eight-step annotation settings. All four workflows follow the same general input and output process: the screenshot is loaded and prepared, the image is passed to one or more LLM prompt branches, the outputs are merged when needed, and the final annotations are converted to JSON and saved. The main difference is the level of task decomposition: the one-step workflow uses a single LLM call for all selected UI element classes, while the two-step, four-step, and eight-step workflows divide the annotation task into two, four, and eight prompt branches, respectively. This consistent structure ensures that performance differences are mainly caused by the level of workflow decomposition rather than by changes in preprocessing or output handling.

\subsection{Evaluation Metrics}

The model-generated annotations are evaluated by comparing them with the expert ground truth annotations from the MUIAnno dataset. The evaluation is performed at the element level. Each predicted annotation is matched with a ground truth annotation when the predicted class is correct and the predicted bounding box sufficiently overlaps with the ground truth bounding box. Correctly matched predictions are counted as true positives. Predictions that do not match any ground truth element are counted as false positives. Ground truth elements that are not detected by the model are counted as false negatives. The study reports precision, recall, and F1-score. Precision measures how many of the model predictions are correct and is calculated as:
\begin{equation}
\label{eq:precision}
Precision = \frac{TP}{TP + FP}
\end{equation}

where $TP$ is the number of true positives and $FP$ is the number of false positives. Precision is the primary metric in this work because the goal is to produce reliable annotations. In an automated annotation pipeline, false positives are costly because they add noise to the dataset and increase the amount of manual correction needed from human reviewers. Recall measures how many ground truth elements are successfully detected by the model and is calculated as:
\begin{equation}
\label{eq:recall}
Recall = \frac{TP}{TP + FN}
\end{equation}

where $FN$ is the number of false negatives. Recall is important because an annotation system should not miss too many valid UI elements. However, high recall alone is not sufficient for this study. A workflow may detect more elements by producing many extra predictions, but this can reduce annotation reliability if many of those predictions are wrong. The F1-score is also reported to summarize the balance between precision and recall and is calculated as:
\begin{equation}
\label{eq:f1}
F1 = \frac{2 \times Precision \times Recall}{Precision + Recall}
\end{equation}

The F1-score is useful for comparing the overall behavior of different workflows, but it is not treated as the main objective. Since the proposed method is intended for model-assisted dataset construction and UI annotation, precision remains the main focus. A high-precision workflow is more useful in practice because it produces annotations that are more trustworthy and require less cleanup \citep{powers2011evaluation}. The evaluation is conducted consistently across all workflow strategies and grouping experiments. The same matching rule, same ground truth annotations, and same metric calculations are used for every experiment. This makes it possible to attribute performance differences to workflow decomposition and class grouping rather than to changes in the evaluation process. The final results are analyzed both overall and by element class to understand which UI components benefit from decomposition and which remain difficult to annotate reliably.

\section{Results and Discussion}
\label{sec:results_discussion}
This section presents the results of the proposed context-aware workflow decomposition approach for automated mobile UI annotation. The analysis focuses on how different workflow structures affect annotation quality when a multimodal large language model is used to generate structured UI element annotations. Before conducting the main workflow decomposition experiments, an initial model selection experiment was performed to compare different multimodal LLMs and select a practical model for the remaining experiments. The main workflow comparison is then conducted across four annotation strategies: one-step, two-step, four-step, and eight-step annotation. In addition, element-level and grouping-level analyses are presented to better understand why some workflow designs perform better than others. The discussion is organized around three main questions. First, does decomposing the annotation task improve the reliability of model-generated UI annotations? Second, how does the level of decomposition affect the balance between precision and recall? Third, do semantically related UI element classes benefit from being annotated together? Since this study is primarily concerned with reliable automated annotation, precision is treated as the most important metric. Recall and F1-score are also reported to show the trade-off between detecting more UI elements and avoiding false positives.

\subsection{Model Selection}

\begin{figure}[H]
\centering
\includegraphics[width=0.78\textwidth]{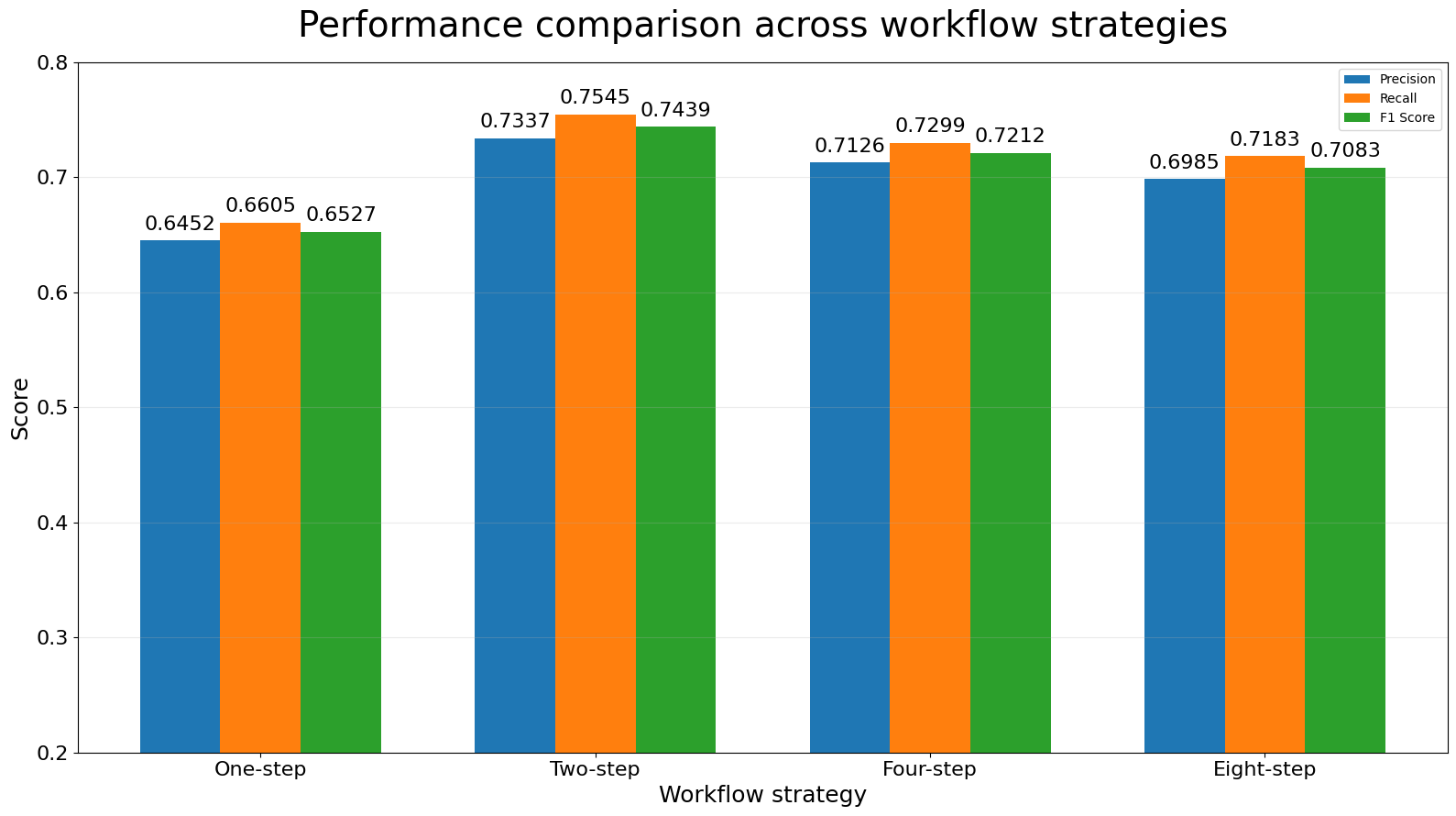}
\caption{Performance comparison of GPT 5.4 across different workflow decomposition strategies.}
\label{fig:openai_model_selection}
\end{figure}

\begin{figure}[H]
\centering
\includegraphics[width=0.78\textwidth]{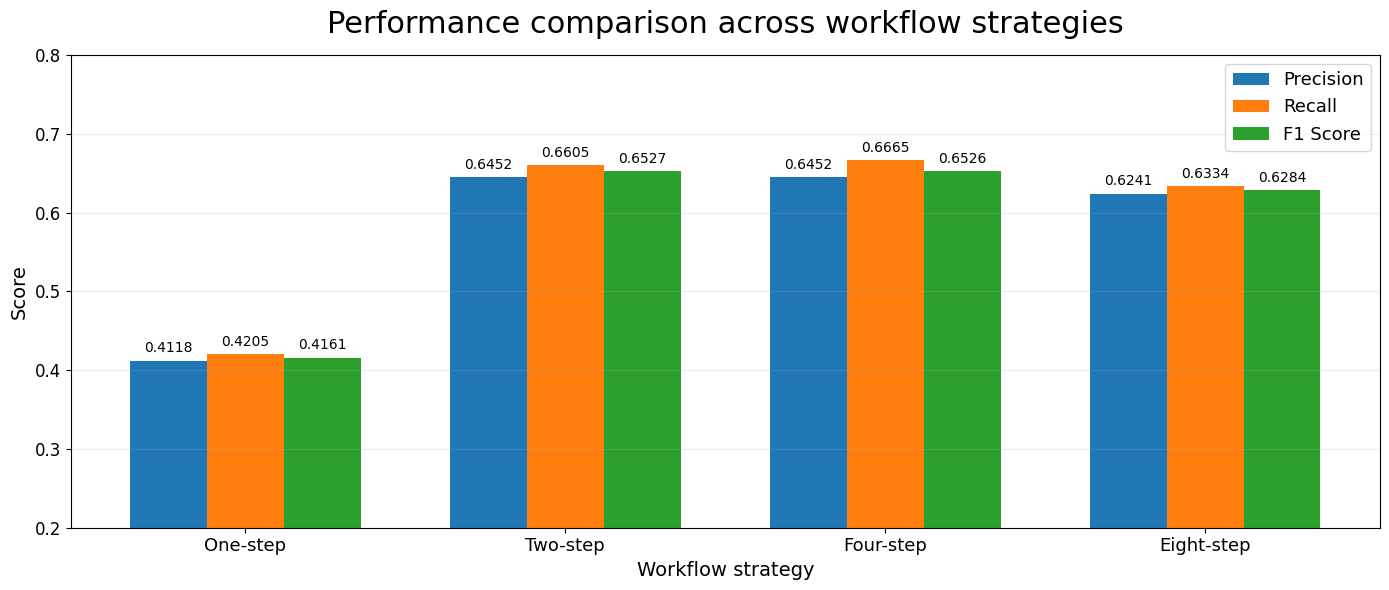}
\caption{Performance comparison of Gemma-4-31B-IT across different workflow decomposition strategies.}
\label{fig:gemma_model_selection}
\end{figure}

\begin{figure}[H]
\centering
\includegraphics[width=0.78\textwidth]{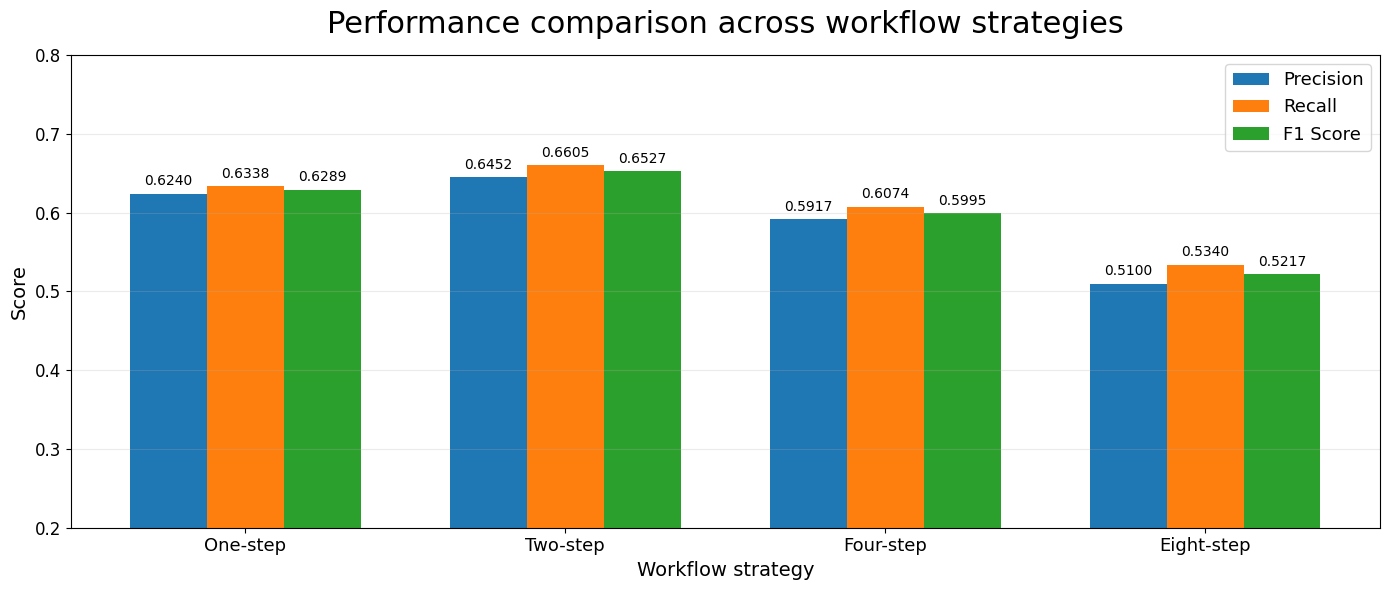}
\caption{Performance comparison of Claude Opus 4.6 across different workflow decomposition strategies.}
\label{fig:claude_model_selection}
\end{figure}

\begin{figure}[H]
\centering
\includegraphics[width=0.78\textwidth]{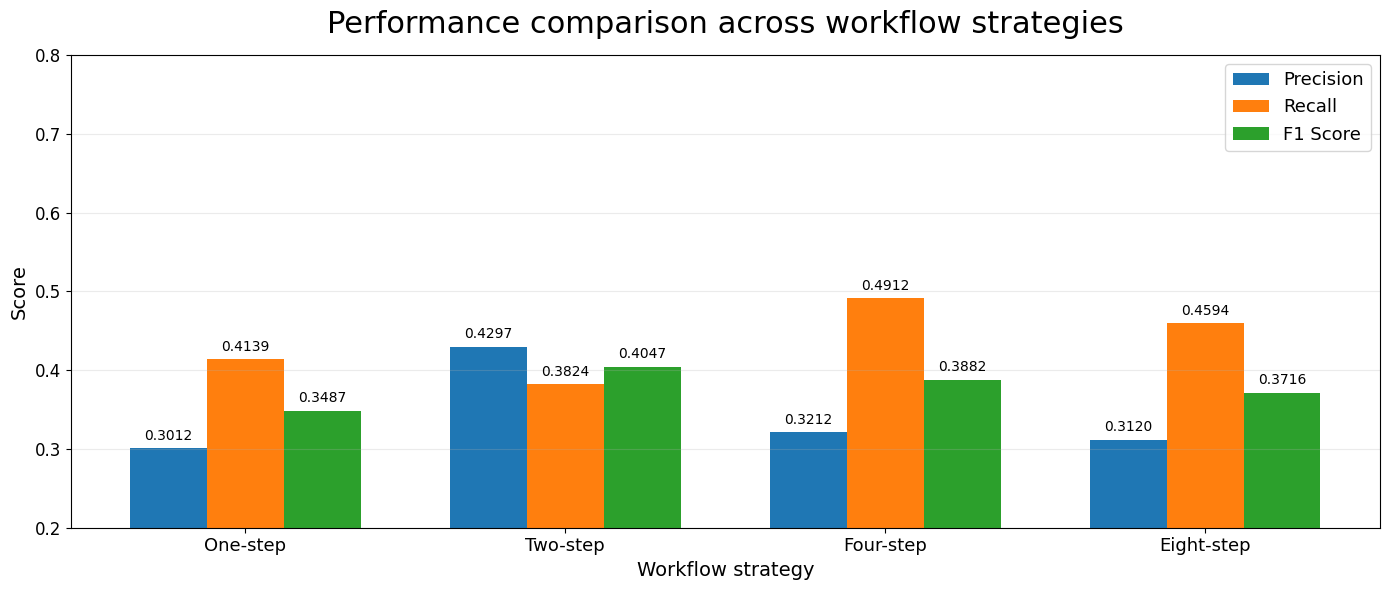}
\caption{Performance comparison of Gemini 3.1 Pro across different workflow decomposition strategies.}
\label{fig:gemini_model_selection}
\end{figure}

Before conducting the main workflow decomposition experiments, an initial model selection experiment was carried out to identify a suitable multimodal LLM for the main annotation pipeline. Four models were compared: GPT 5.4 \citep{openai2026gpt54}, Gemma-4-31B-IT \citep{googledeepmind2026gemma4}, Claude Opus 4.6 \citep{anthropic2026claudeopus46}, and Gemini 3.1 Pro \citep{googledeepmind2026gemini31pro}. The purpose of this experiment was not to present a full benchmark of commercial and open multimodal models. Instead, it was used as a practical screening step to select a model that could support the remaining workflow experiments with an acceptable balance between annotation quality, consistency, and deployment cost. All four models were evaluated under the same experimental conditions. The same set of mobile UI screenshots, selected UI element classes, workflow configurations, prompt structure, output schema, and evaluation metrics were used for each model. This ensured that the comparison focused on model behavior rather than differences in data preparation or evaluation procedure. Each model was tested using the one-step, two-step, four-step, and eight-step workflow settings. The outputs were parsed as structured JSON annotations and evaluated against the expert ground truth using precision, recall, and F1-score. Since this study emphasizes reliable annotation, the best precision-oriented workflow for each model was used in the model selection summary. Table~\ref{tab:model_selection} presents the initial model selection results. GPT 5.4 achieved the strongest overall performance among the compared models. Its best result was obtained using the two-step workflow, which produced the highest precision, recall, and F1-score in the model selection experiment. This indicates that GPT 5.4 was the most accurate model in terms of both detecting UI elements and avoiding incorrect predictions. Gemma-4-31B-IT and Claude Opus 4.6 also showed competitive results, with both models performing best under the two-step workflow. Their scores suggest that they were able to benefit from moderate workflow decomposition in a similar way. Gemini 3.1 Pro produced lower absolute scores than GPT 5.4, Gemma-4-31B-IT, and Claude Opus 4.6. However, it was selected for the main workflow experiments because it provided a more cost-efficient and scalable setting for repeated annotation runs. This decision was important because the main objective of the paper is not to compare model providers, but to study how workflow decomposition and class grouping affect automated UI annotation quality. The main experiments required multiple workflow runs, element-level evaluations, and combinational grouping experiments. Therefore, using a model with lower operational cost made it possible to evaluate the proposed workflow design more extensively while keeping the experimental setup practical.\begin{table}[h]
\centering
\caption{Initial model selection results using the best precision workflow for each model.}
\label{tab:model_selection}
\resizebox{\textwidth}{!}{%
\begin{tabular}{l c c c c p{4.5cm}}
\hline
\textbf{Model} & \textbf{Selected Workflow} & \textbf{Precision} & \textbf{Recall} & \textbf{F1-score} & \textbf{Decision} \\
\hline
GPT 5.4 & Two-step & 0.734 & 0.754 & 0.744 & Highest overall performance, but higher cost \\
Gemma-4-31B-IT & Two-step & 0.645 & 0.660 & 0.653 & Competitive performance, but not selected \\
Claude Opus 4.6 & Two-step & 0.645 & 0.660 & 0.653 & Competitive performance, but not selected \\
Gemini 3.1 Pro & Two-step & 0.430 & 0.382 & 0.405 & Selected for the main experiments due to cost efficiency \\
\hline
\end{tabular}%
}
\end{table} An important observation from the model selection experiment is that the two-step workflow consistently appeared as the strongest setting for all four models. This pattern suggests that the benefit of moderate decomposition is not limited to one model. When all UI classes are annotated together in a single prompt, the task becomes broad and more difficult to control. When each class is annotated separately, the model may lose useful context for distinguishing visually similar elements. The two-step setting provides a middle ground by reducing prompt complexity while still preserving enough contextual information for the model to compare related UI components. To provide a clearer visual comparison of model behavior across workflow strategies, Figures~\ref{fig:openai_model_selection}--\ref{fig:gemini_model_selection} present the precision, recall, and F1-score results for each model separately. These figures complement Table~\ref{tab:model_selection} by showing how performance changes across the one-step, two-step, four-step, and eight-step workflows.

As shown in Figures~\ref{fig:openai_model_selection}--\ref{fig:gemini_model_selection}, GPT 5.4 achieved the best overall results, while Gemma-4-31B-IT and Claude Opus 4.6 showed similar and competitive behavior. Gemini 3.1 Pro had lower absolute performance, but its results were stable enough to support the main purpose of the study. For this reason, all subsequent workflow decomposition, element-level, and grouping experiments were conducted using Gemini 3.1 Pro as the base multimodal model. This choice allowed the study to focus on the central research question: how workflow decomposition and UI class grouping affect the reliability of automated mobile UI annotation.

\subsection{Overall Workflow Comparison}

Table~\ref{tab:overall_workflow_results} presents the overall performance of the four evaluated workflow strategies using Gemini 3.1 Pro. The one-step workflow asks the model to annotate all eight UI element classes in a single prompt. The two-step workflow divides the eight classes into two broader groups. The four-step workflow annotates two classes per stage, while the eight-step workflow annotates one class at a time. \begin{table}[h]
\centering
\caption{Overall performance comparison of annotation workflows using Gemini 3.1 Pro.}
\label{tab:overall_workflow_results}
\resizebox{\textwidth}{!}{%
\begin{tabular}{l p{6.2cm} c c c}
\hline
\textbf{Workflow} & \textbf{Description} & \textbf{Precision} & \textbf{Recall} & \textbf{F1-score} \\
\hline
One-step & All eight elements annotated together in a single workflow stage & 0.3012 & 0.4139 & 0.3487 \\
Two-step & Four elements annotated first, followed by the remaining four elements & 0.4297 & 0.3824 & 0.4047 \\
Four-step & Two elements annotated per workflow stage & 0.3212 & 0.5112 & 0.3945 \\
Eight-step & One element annotated per workflow stage & 0.3120 & 0.5694 & 0.4031 \\
\hline
\end{tabular}%
}
\end{table} The results show that workflow design has a clear effect on annotation quality. The two-step workflow achieves the highest precision, with a score of 0.4297. This is the most important result for this study because precision reflects the reliability of the generated annotations. In practical annotation pipelines, a high number of false positives can reduce dataset quality and increase the amount of manual correction required. Therefore, the two-step workflow is more useful than workflows that detect more elements but introduce more incorrect predictions. The two-step workflow also achieves the highest F1-score, with a value of 0.4047. Compared with the one-step workflow, this represents an improvement of approximately 16.1\% in F1-score. This improvement suggests that asking the model to annotate all eight UI classes at once creates unnecessary prompt complexity. In the one-step setting, the model must simultaneously distinguish between interactive, textual, visual, and structural elements. This increases ambiguity and leads to more incorrect predictions. The four-step and eight-step workflows show a different pattern. Both achieve higher recall than the two-step workflow, with the eight-step workflow reaching the highest recall of 0.5694. This means that deeper decomposition helps the model detect more ground truth elements. However, this improvement in recall comes at the cost of precision. The eight-step workflow has a precision of only 0.3120, which is lower than the two-step workflow. This indicates that annotating one class at a time encourages the model to predict more candidate elements, but many of these predictions are not correct. This result is important because it shows that more decomposition is not automatically better. A highly decomposed workflow reduces the number of target classes in each prompt, but it also removes useful context. Many UI elements are defined not only by their appearance but also by their relationship to surrounding elements. When each class is annotated independently, the model may lose the ability to compare similar classes within the same reasoning step. For example, clickable text may be confused with plain text, and tabs may be confused with buttons or labels. Therefore, the best workflow is not the deepest one, but the one that balances task simplicity with enough contextual information. Overall, the workflow-level results support the main assumption of this paper: annotation performance depends not only on the multimodal model itself, but also on how the annotation task is organized. The two-step workflow provides the strongest balance because it reduces prompt complexity while still preserving enough screen-level context for the model to distinguish related UI components. In contrast, excessive decomposition increases recall but also introduces more false positives, reducing the reliability of the generated annotations.

\subsection{Two-Step Workflow Analysis}

The overall workflow results show that the two-step workflow achieves the highest precision and F1-score among the evaluated Gemini 3.1 Pro workflows. This result suggests that moderate decomposition is more effective than both annotating all classes together and annotating each class separately. The main reason is that the two-step workflow reduces the complexity of the prompt while still keeping enough contextual information for the model to compare related UI elements. In the one-step workflow, the model is asked to annotate all eight UI element classes in a single prompt. This gives the model full access to the screen context, but it also makes the task too broad. The model has to identify interactive elements, textual elements, visual elements, and structural containers at the same time. As a result, visually similar classes can easily be confused. For example, a tab may be predicted as a button, clickable text may be confused with plain text, and a card may be missed or incorrectly detected because its boundary is not always visually clear. This broad setting increases the chance of false positives and reduces annotation reliability. The eight-step workflow has the opposite problem. In this setting, each UI element class is annotated independently. This makes each prompt simpler, but it also removes useful relationships between related classes. Many UI elements cannot be understood correctly in isolation. A short text region, for example, may be a label, plain text, clickable text, or tab depending on its position and surrounding layout. Similarly, an icon may appear as a standalone action, inside a button, inside a tab, or as a decorative visual element. When the model sees only one target class at a time, it may over-predict that class because it is not directly comparing it with similar alternatives. This explains why the eight-step workflow achieves higher recall but lower precision.

The element-level results also support this interpretation. Classes such as plain text and label are relatively easier for the model because they usually contain readable textual content and clearer semantic cues. Button and image show moderate performance because they often have more visible boundaries or distinctive visual regions. In contrast, clickable text, tab, card, and icon are more difficult because their correct interpretation depends strongly on surrounding context. Clickable text may look almost identical to ordinary text, tabs may resemble buttons or labels, cards may be defined by layout grouping rather than clear borders, and icons are often small or nested inside other components. The two-step workflow provides a better balance between these two extremes. It reduces the number of element classes that the model must handle in each prompt, which makes the annotation task more focused than the one-step workflow. At the same time, it does not separate every class completely, so the model can still use contextual relationships between related UI elements. This balance helps reduce unnecessary predictions while preserving enough semantic context for distinguishing similar classes. Therefore, the stronger performance of the two-step workflow is not caused by decomposition alone. It is mainly due to the balance between prompt simplicity and context preservation. The results indicate that automated UI annotation should not simply use the smallest possible prompt or the largest possible prompt. Instead, the workflow should group UI element classes in a way that keeps related elements together while avoiding unnecessary prompt complexity. A more detailed combinational grouping analysis is presented in Section~\ref{sec:combinational_grouping_analysis}.

\subsection{Element-Level Performance}

To understand why the workflow structure affects performance, each of the eight selected UI element classes was first evaluated independently. This single-element setting represents the most decomposed case, where the model focuses on only one UI class at a time. The purpose of this analysis is to identify which elements can be detected reliably in isolation and which elements need surrounding context to be annotated correctly. Table~\ref{tab:single_element_results} presents the results. \begin{table}[h]
\centering
\caption{Single-element annotation performance.}
\label{tab:single_element_results}
\small
\begin{tabularx}{0.98\linewidth}{p{1.4cm} p{2.3cm} c c c X}
\hline
\textbf{Workflow} & \textbf{Element} & \textbf{Precision} & \textbf{Recall} & \textbf{F1-score} & \textbf{Observation} \\
\hline
1-1 & Button & 0.4232 & 0.4618 & 0.4416 & Moderate performance \\
1-2 & Tab & 0.1574 & 0.8611 & 0.2661 & High recall but very low precision \\
1-3 & Clickable Text & 0.1222 & 0.7463 & 0.2101 & Many false positives \\
1-4 & Card & 0.1745 & 0.6981 & 0.2792 & Many false positives \\
1-5 & Label & 0.5158 & 0.4030 & 0.4524 & Relatively better precision \\
1-6 & Plain Text & 0.5446 & 0.3179 & 0.4015 & Best precision among single-element workflows \\
1-7 & Icon & 0.1971 & 0.3006 & 0.2380 & Difficult element type for the model \\
1-8 & Image & 0.3613 & 0.7667 & 0.4911 & Best single-element F1-score \\
\hline
\end{tabularx}
\end{table}The results show that the model does not annotate all UI classes with the same level of reliability. \textit{Plain text} achieves the highest single-element precision, with a value of 0.5446, followed by \textit{label} with 0.5158. These two classes are easier for the model because they usually contain readable textual content and have clearer visual cues. The \textit{button} class also performs moderately well, with a precision of 0.4232, because many buttons have visible boundaries, alignment patterns, or action-oriented text. The \textit{image} class achieves the highest single-element F1-score of 0.4911. This suggests that images are often easier to detect when the model focuses only on that class, probably because image regions tend to occupy larger areas and have stronger visual separation from surrounding text. However, the precision of the image class is still limited, which means that the model may sometimes confuse thumbnails, icons, decorative graphics, or background visual regions with valid image elements. In contrast, \textit{clickable text}, \textit{tab}, \textit{card}, and \textit{icon} are much harder to annotate independently. Clickable text has the lowest precision, with a score of 0.1222. This is expected because clickable text often looks very similar to plain text or labels. Unless there is a strong visual cue such as color, underline, placement, or navigation context, the model may incorrectly mark ordinary text as clickable. The \textit{tab} class also has very low precision, 0.1574, despite achieving high recall. This means that the model detects many possible tab-like regions, but most of these predictions are false positives. Tabs may be confused with buttons, labels, navigation text, or segmented controls. The \textit{card} class is also difficult because cards are often defined by spacing, grouping, shadows, or background regions rather than by clear borders. Similarly, icons are small, visually diverse, and often nested inside buttons, cards, or navigation bars. These results show that some UI classes cannot be reliably understood in isolation. Their correct interpretation depends on nearby text, layout structure, and functional context. This finding helps explain why the two-step workflow performs better than the eight-step workflow: it keeps related classes together, giving the model more context for comparison.

\subsection{Two-Element Grouping}

The two-element workflow analysis examines whether grouping related UI element types improves annotation behavior compared with annotating each element independently. Table~\ref{tab:two_element_results} shows the results for four two-element groups. \begin{table}[h]
\centering
\caption{Two-element workflow performance.}
\label{tab:two_element_results}
\small
\begin{tabularx}{0.98\linewidth}{p{1.5cm} X c c c X}
\hline
\textbf{Workflow} & \textbf{Elements} & \textbf{Precision} & \textbf{Recall} & \textbf{F1-score} & \textbf{Observation} \\
\hline
2-1 & Button, Tab & 0.3202 & 0.5612 & 0.4077 & Better balance than Tab alone \\
2-2 & Clickable Text, Card & 0.1974 & 0.7583 & 0.3133 & High recall but low precision \\
2-3 & Label, Plain Text & 0.5501 & 0.4430 & 0.4908 & Best two-element group \\
2-4 & Icon, Image & 0.2434 & 0.3901 & 0.2997 & Visual element grouping remains difficult \\
\hline
\end{tabularx}
\end{table}The best two-element group is \textit{label} and \textit{plain text}, with a precision of 0.5501 and an F1-score of 0.4908. This confirms that text-based elements are among the most reliable classes for the model. More importantly, it shows that grouping related text classes can help the model compare their roles within the same prompt. A label and plain text may look visually similar, but their function in the interface is different. When both are included together, the model has a better chance of separating semantic text from ordinary readable content. The \textit{button} and \textit{tab} group also improves the balance compared with annotating tabs alone. In the single-element setting, the tab class had high recall but very low precision. When tabs are grouped with buttons, the model can compare two related interactive categories, which appears to reduce some of the false positives. However, the precision remains moderate, showing that interactive navigation elements are still difficult when their visual boundaries are subtle. The \textit{clickable text} and \textit{card} group achieves high recall but low precision. This means that the model detects many candidate regions but also introduces many incorrect predictions. Both classes are highly context-dependent, but they depend on different kinds of context. Clickable text depends on functional and semantic cues, while cards depend more on layout grouping and visual structure. Grouping them together does not fully resolve the ambiguity. The \textit{icon} and \textit{image} group also remains difficult. Although both are visual elements, they differ in size, purpose, and boundary clarity. Icons are usually small and symbolic, while images are usually larger content regions. The model may still confuse decorative graphics, content images, thumbnails, and symbolic icons. These results show that grouping is useful only when the grouped classes provide meaningful comparison. Textual grouping helps because the classes are related but separable. In contrast, grouping two visually ambiguous classes does not automatically improve precision.

\subsection{Four and Eight Element Grouping}

The four-element and eight-element grouped workflows provide a broader view of how class grouping affects annotation performance. Table~\ref{tab:grouped_workflow_results} presents the results. \begin{table}[h]
\centering
\caption{Grouped workflow performance.}
\label{tab:grouped_workflow_results}
\small
\begin{tabularx}{0.98\linewidth}{p{1.7cm} X c c c X}
\hline
\textbf{Workflow} & \textbf{Elements} & \textbf{Precision} & \textbf{Recall} & \textbf{F1-score} & \textbf{Observation} \\
\hline
4-1 & Button, Tab, Clickable Text, Card & 0.2857 & 0.4798 & 0.3582 & Interactive and structural elements are harder \\
4-2 & Label, Plain Text, Icon, Image & 0.5337 & 0.2649 & 0.3540 & Higher precision but lower recall \\
8-elements & All eight elements & 0.3112 & 0.4229 & 0.3585 & Mixed element types increase ambiguity \\
\hline
\end{tabularx}
\end{table}The results show that different groups produce different error patterns. The group containing \textit{button}, \textit{tab}, \textit{clickable text}, and \textit{card} achieves higher recall than the group containing \textit{label}, \textit{plain text}, \textit{icon}, and \textit{image}. However, its precision is much lower. This indicates that interactive and structural elements encourage the model to produce more predictions, but many of those predictions are incorrect. These classes are harder to separate because they often overlap visually or functionally. For example, a tab may look like a button, clickable text may look like a label, and a card may contain text, buttons, icons, and images. The second group, containing \textit{label}, \textit{plain text}, \textit{icon}, and \textit{image}, achieves much higher precision of 0.5337 but lower recall of 0.2649. This means that when the model predicts elements in this group, the predictions are more likely to be correct, but the model misses more ground truth elements. For a precision-oriented annotation pipeline, this behavior is useful because it produces cleaner annotations. However, it may not be sufficient when high coverage is also required. The eight-element workflow has an F1-score of 0.3585, which is close to the four-element groups but lower than the two-step workflow reported in Table~\ref{tab:overall_workflow_results}. This supports the idea that putting all classes into a single prompt increases ambiguity. Although the model receives the full screen context, the task becomes too broad because the model must distinguish many visually similar and nested elements at once. As a result, the benefit of full context is reduced by the complexity of the prompt.

\subsection{Combinational Grouping Analysis}
\label{sec:combinational_grouping_analysis}
The previous results show that the two-step workflow performs best overall, but they do not fully explain whether this improvement comes only from reducing the number of stages or from the way UI classes are grouped. To examine this more carefully, an additional combinational grouping analysis was conducted. The analysis used the same eight frequent UI element classes: \textit{button}, \textit{tab}, \textit{clickable text}, \textit{card}, \textit{label}, \textit{plain text}, \textit{icon}, and \textit{image}. All possible four-element combinations were generated from these eight classes:

\[
\binom{8}{4} = 70
\]

Since each four-element group has a complementary four-element group, these 70 evaluated groups correspond to 35 unique 4--4 workflow splits. Each group was evaluated using the same prompt structure, output schema, model setting, and evaluation protocol. This analysis is important because it tests whether the two-step workflow works well simply because it has two stages, or because certain class combinations preserve useful context. Table~\ref{tab:single_vs_grouped_precision} compares the single-element precision with the best grouped precision for selected classes. The largest improvement is observed for \textit{clickable text}. Its precision increases from 0.1222 in the single-element setting to 0.2432 when grouped with related elements. Although the grouped precision is still modest, the relative improvement is important because clickable text is the most difficult class in the isolated setting. This shows that clickable text benefits from being interpreted alongside other text-related elements. \begin{table}[h]
\centering
\caption{Comparison between single-element precision and best grouped precision.}
\label{tab:single_vs_grouped_precision}
\small
\begin{tabularx}{0.98\linewidth}{p{2.7cm} c c X}
\hline
\textbf{UI Element} & \textbf{Single P} & \textbf{Grouped P} & \textbf{Observation} \\
\hline
Plain Text & 0.5446 & 0.5612 & Slight improvement with contextual grouping \\
Label & 0.5158 & 0.5117 & Comparable performance, indicating stable text recognition \\
Clickable Text & 0.1222 & 0.2432 & Clear improvement because grouping helps distinguish clickable text from ordinary text \\
Image & 0.3613 & 0.3841 & Improved precision due to related visual-text context \\
\hline
\end{tabularx}
\end{table}The \textit{plain text} class also improves slightly under grouping, increasing from 0.5446 to 0.5612. The \textit{image} class improves from 0.3613 to 0.3841, suggesting that nearby textual context can help the model decide whether a visual region should be treated as an image. 
The \textit{label} class remains almost unchanged, decreasing only slightly from 0.5158 to 0.5117. This suggests that labels already contain strong semantic cues and do not depend on grouping as much as clickable text. Among all evaluated combinations, the best observed four-class group was \textit{Clickable Text, Label, Plain Text, and Image}. This group achieved the highest grouped precision of 0.4544. For the same four classes, the average single-element precision was 0.3860. This means that meaningful grouping improved precision by preserving useful context between related text and image elements. 
Table~\ref{tab:best_group_combination} summarizes this comparison.\begin{table}[h]
\centering
\caption{Best observed four-class grouping compared with the single-element baseline.}
\label{tab:best_group_combination}
\small
\begin{tabularx}{0.98\linewidth}{X c X}
\hline
\textbf{Setting} & \textbf{Precision} & \textbf{Interpretation} \\
\hline
Average single-element precision of Clickable Text, Label, Plain Text, and Image & 0.3860 & Elements are annotated independently, so the model has limited class-level context \\
Best grouped precision for Clickable Text + Label + Plain Text + Image & 0.4544 & Grouping related text and image elements preserves context and reduces ambiguity \\
\hline
\end{tabularx}
\end{table}This result explains why excessive decomposition is less effective. When each class is annotated independently, the model loses the opportunity to compare related classes. This is especially harmful for classes whose meaning depends on context. A short text region, for example, may be a label, plain text, clickable text, or tab depending on its location and surrounding layout. Similarly, an icon may be a standalone control, part of a button, part of a tab, or a decorative symbol. Without related classes in the same prompt, the model may over-predict the target class and increase false positives. Overall, the combinational analysis shows that the success of the two-step workflow is not only due to reducing the number of stages. It is also due to preserving useful relationships between UI elements. The one-step workflow keeps all context but makes the prompt too broad. The eight-step workflow makes the prompt simple but removes too much context. The two-step workflow works better because it sits between these two extremes.

\subsection{Evaluated Group Combinations}

Table~\ref{tab:all_70_combinations} lists the 70 four-element UI class combinations evaluated in the combinational grouping analysis. These combinations were generated from the eight selected UI element classes: button, tab, clickable text, card, label, plain text, icon, and image. The results show that context-aware workflow decomposition can improve automated UI annotation, but only when decomposition is used carefully. The one-step workflow is too broad because it asks the model to annotate all eight UI classes at once. This increases prompt complexity and makes the model more likely to confuse visually similar elements. The eight-step workflow is too narrow because it separates every class into an independent prompt. Although this increases recall, it also removes useful class-level context and produces more false positives. The two-step workflow provides the best balance. It reduces the complexity of the annotation task compared with the one-step workflow, but it still keeps related UI elements together. This is important because mobile UI elements are not independent objects. Their meaning often depends on surrounding components, layout structure, visual grouping, and interaction patterns. A text region, for example, can only be interpreted correctly when the model understands whether it appears inside a button, below an image, inside a card, or as part of a navigation area. The element-level results show that some classes are naturally easier for the model to annotate. Plain text, labels, and images perform relatively well because they have clearer visual cues. In contrast, clickable text, tabs, cards, and icons are more difficult because they are smaller, more ambiguous, or more dependent on surrounding context. These are also the classes most likely to produce false positives. This suggests that future annotation systems should not rely on one universal prompt for all UI elements. Instead, they should use class-aware or group-aware prompts that reflect the visual and semantic relationships between UI components. The grouped and combinational experiments provide the clearest explanation for the strong performance of the two-step workflow. Grouping should not be based only on equal class counts. It should also consider semantic compatibility. The best group, \textit{Clickable Text + Label + Plain Text + Image}, shows that text-heavy and visual-content elements can support each other during annotation. In contrast, groups dominated by highly ambiguous interactive or structural elements still produce lower precision. Therefore, workflow design should consider how UI classes relate visually and functionally. From a practical perspective, these findings are important for automated dataset construction.

\begin{table}[H]
\centering
\caption{All 70 evaluated four-element UI class combinations.}
\label{tab:all_70_combinations}
\footnotesize
\renewcommand{\arraystretch}{1.0}
\begin{tabular}{p{0.32\textwidth} p{0.32\textwidth} p{0.30\textwidth}}
\hline
\textbf{Combinations 1--24} & \textbf{Combinations 25--48} & \textbf{Combinations 49--70} \\
\hline
1. Button, Tab, Clickable Text, Label &
25. Button, Tab, Plain Text, Image &
49. Button, Card, Label, Plain Text \\

2. Card, Plain Text, Icon, Image &
26. Clickable Text, Card, Label, Icon &
50. Tab, Clickable Text, Icon, Image \\

3. Button, Tab, Clickable Text, Plain Text &
27. Button, Tab, Icon, Image &
51. Button, Card, Label, Icon \\

4. Card, Label, Icon, Image &
28. Clickable Text, Card, Label, Plain Text &
52. Tab, Clickable Text, Plain Text, Image \\

5. Button, Tab, Clickable Text, Icon &
29. Button, Clickable Text, Card, Label &
53. Button, Card, Label, Image \\

6. Card, Label, Plain Text, Image &
30. Tab, Plain Text, Icon, Image &
54. Tab, Clickable Text, Plain Text, Icon \\

7. Button, Tab, Clickable Text, Image &
31. Button, Clickable Text, Card, Plain Text &
55. Button, Card, Plain Text, Icon \\

8. Card, Label, Plain Text, Icon &
32. Tab, Label, Icon, Image &
56. Tab, Clickable Text, Label, Image \\

9. Button, Tab, Card, Label &
33. Button, Clickable Text, Card, Icon &
57. Button, Card, Plain Text, Image \\

10. Clickable Text, Plain Text, Icon, Image &
34. Tab, Label, Plain Text, Image &
58. Tab, Clickable Text, Label, Icon \\

11. Button, Tab, Card, Plain Text &
35. Button, Clickable Text, Card, Image &
59. Button, Card, Icon, Image \\

12. Clickable Text, Label, Icon, Image &
36. Tab, Label, Plain Text, Icon &
60. Tab, Clickable Text, Label, Plain Text \\

13. Button, Tab, Card, Icon &
37. Button, Clickable Text, Label, Plain Text &
61. Button, Label, Plain Text, Icon \\

14. Clickable Text, Label, Plain Text, Image &
38. Tab, Card, Icon, Image &
62. Tab, Clickable Text, Card, Image \\

15. Button, Tab, Card, Image &
39. Button, Clickable Text, Label, Icon &
63. Button, Label, Plain Text, Image \\

16. Clickable Text, Label, Plain Text, Icon &
40. Tab, Card, Plain Text, Image &
64. Tab, Clickable Text, Card, Icon \\

17. Button, Tab, Label, Plain Text &
41. Button, Clickable Text, Label, Image &
65. Button, Label, Icon, Image \\

18. Clickable Text, Card, Icon, Image &
42. Tab, Card, Plain Text, Icon &
66. Tab, Clickable Text, Card, Plain Text \\

19. Button, Tab, Label, Icon &
43. Button, Clickable Text, Plain Text, Icon &
67. Button, Plain Text, Icon, Image \\

20. Clickable Text, Card, Plain Text, Image &
44. Tab, Card, Label, Image &
68. Tab, Clickable Text, Card, Label \\

21. Button, Tab, Label, Image &
45. Button, Clickable Text, Plain Text, Image &
69. Button, Tab, Clickable Text, Card \\

22. Clickable Text, Card, Plain Text, Icon &
46. Tab, Card, Label, Icon &
70. Label, Plain Text, Icon, Image \\

23. Button, Tab, Plain Text, Icon &
47. Button, Clickable Text, Icon, Image &
 \\

24. Clickable Text, Card, Label, Image &
48. Tab, Card, Label, Plain Text &
 \\
\hline
\end{tabular}
\end{table}

In model-assisted annotation, false positives are costly because they must be reviewed and removed by humans. A workflow that produces slightly fewer elements but with higher correctness may be more useful than one that produces many noisy annotations. For this reason, the two-step workflow is the most suitable strategy in this study. It improves precision and F1-score while avoiding the excessive false-positive behavior observed in deeper decomposition. The results also suggest that future systems could use adaptive workflow decomposition. Instead of using a fixed one-step or eight-step strategy, an annotation pipeline could choose groups based on screen complexity, element density, or class ambiguity. For simple screens, a broader prompt may be sufficient. For dense screens with many text elements, a text-focused group may improve precision. For screens with many icons and navigation elements, specialized prompts or post-processing may be needed. This direction could make LLM-based UI annotation more reliable and scalable. Overall, the results confirm that context-aware workflow decomposition is a useful strategy for improving automated mobile UI annotation. Moderate decomposition reduces prompt complexity while preserving enough interface context for semantic reasoning. In contrast, excessive decomposition increases recall but weakens reliability by producing more false positives. These findings show that workflow design is a critical factor in building practical LLM-based annotation pipelines for mobile UI understanding.

\section{Conclusion}
\label{sec:conclusion}

This paper presented a context-aware workflow decomposition approach for automated mobile UI annotation using multimodal large language models. The study first compared multiple multimodal LLMs to select a practical model setting for the main experiments. Although GPT 5.4 achieved the strongest overall performance in the initial comparison, Gemini 3.1 Pro was selected for the main workflow experiments because it provided a more cost-efficient setting for scalable annotation. Using Gemini 3.1 Pro, the study evaluated one-step, two-step, four-step, and eight-step workflows on eight frequent UI element classes from the MUIAnno dataset: \textit{button}, \textit{tab}, \textit{clickable text}, \textit{card}, \textit{label}, \textit{plain text}, \textit{icon}, and \textit{image}. The results show that the two-step workflow achieved the best overall performance, especially in precision. This indicates that moderate task decomposition can reduce prompt complexity while still preserving useful screen context. The findings also show that UI annotation quality depends not only on the model, but also on how UI element classes are grouped. Single-element results showed that classes such as \textit{plain text}, \textit{label}, and \textit{image} were easier to annotate, while \textit{clickable text}, \textit{tab}, \textit{card}, and \textit{icon} remained more difficult because they are often small, nested, visually ambiguous, or context-dependent. The combinational grouping analysis further showed that meaningful grouping can improve precision. Among the evaluated four-element combinations, the best group was \textit{Clickable Text + Label + Plain Text + Image}, which achieved higher precision than the average isolated precision of the same elements. Overall, the study shows that workflow design is an important factor in reliable LLM-based UI annotation. A two-step workflow provides a practical balance between reducing prompt complexity and preserving contextual relationships between UI elements. Future work can extend this approach to the full UI taxonomy, larger datasets, additional multimodal models, and adaptive grouping strategies for more complex mobile screens.

\section*{Disclosure Statement}

The author reports no potential conflict of interest.

\section*{Funding}

No funding was received for this work.

\section*{Data Availability Statement}

The MUIAnno dataset is publicly available at: \url{https://huggingface.co/datasets/atharparvezce/MUIAnno-Mobile-UI-Dataset}. The annotation tool is publicly available at: \url{https://annota-nine.vercel.app}.

\bibliographystyle{apalike}
\bibliography{references}

\end{document}